\documentclass[12pt]{article}

\catcode`\@=11

\global\arraycolsep=2pt
\usepackage{amsbsy,amssymb,latexsym,amsfonts,amsmath}

\begin{document}
\begin{flushright}
\parbox{4.2cm}
{RUP-16-22}
\end{flushright}

\vspace*{0.7cm}

\begin{center}
{ \Large Euclidean M-theory background dual to three-dimensional scale invariant field theory without conformal invariance}
\vspace*{1.5cm}\\
{Yu Nakayama}
\end{center}
\vspace*{1.0cm}
\begin{center}

Department of Physics, Rikkyo University, Toshima, Tokyo 171-8501, Japan

\vspace{3.8cm}
\end{center}

\begin{abstract}
We show that  eleven dimensional supergravity in Euclidean signature admits an exact classical solution with isometry corresponding to a three dimensional scale invariant field theory without conformal invariance. We also construct the holographic renormalization group flow that connects the known UV conformal fixed point  and the new scale invariant but not conformal fixed point. In view of holography, the existence of such classical solutions suggests that the topologically twisted M2-brane gauge theory possesses a scale invariant but not conformal phase.

\end{abstract}

\thispagestyle{empty} 

\setcounter{page}{0}

\newpage

The enhancement of conformal invariance from scale invariance in quantum field theories  at first sight is quite surprising but careful studies reveal its ubiquity \cite{Nakayama:2013is}. In particular, in two \cite{Polchinski:1987dy} and four dimensions \cite{Luty:2012ww}\cite{Dymarsky:2013pqa}, we have culminating evidence that it is indeed the case.\footnote{See also \cite{Nakayama:2011tk}\cite{Fortin:2012hn}\cite{Grinstein:2013cka}\cite{Bzowski:2014qja}\cite{Yonekura:2014tha}\cite{Rosten:2014oja}\cite{Dymarsky:2015jia}\cite{Dymarsky:2015jia}\cite{Sachs:2015yba}\cite{Bianchi:2016viy}\cite{Naseh:2016maw} for recent discussions.} The situation in other dimensions are less clear. Furthermore, all the argument so far relies on Lorentz invariance in crucial ways so that the enhancement may not happen in Euclidean systems.

Indeed, in our recent paper \cite{Nakayama:2016ydc}, we have pointed out that Euclidean field theories admit more general deformations than usually discussed in quantum field theories because of mixing between rotational symmetry and internal symmetry (a.k.a topological twist \cite{Witten:1988ze}\cite{Witten:1988xj}). Such deformations may be relevant, and if the subsequent renormalization group flow leads to a non-trivial fixed point, it generically gives rise to a scale invariant Euclidean field theory without conformal invariance.
We have further showed that the Euclidean Einstein-Yang-Mills system in four dimensions possesses the holographic solution corresponding to such a fixed point.

Let us consider the Euclidean Einstein-Yang-Mills theory in four dimensions
\begin{align}
 S = \int d^4x \sqrt{g} \left( R +24m^2 - G^a_{\mu\nu}G^{a\mu\nu} \right) .
\end{align}
The Yang-Mills part of the solution is given by the $\phi^4$ theory ansatz for the $SU(2)$ Yang-Mills equation in four dimensions \cite{Corrigan:1976vj}\cite{Actor:1979in} (we use the index notation $i=1,2,3$, $a= 1,2,3$ and $\mu=z,1,2,3$)
\begin{align}
& \partial_\mu \partial^\mu \phi + \lambda \phi^3 = 0  \ \label{phi} \cr
& g_{\mathrm{YM}} A_z^a = \mp \delta^a_i \frac{{\partial^i} \phi}{\phi} \ , \ \ g_{\mathrm{YM}} A_i^a = \epsilon_{i\ k}^{\ a} \partial^k \phi \pm \delta_{i}^a\frac{\partial_z \phi}{\phi} \ ,
\end{align}
where $\lambda$ is arbitrary and $\epsilon_{i\ k}^{\ a}$ is $SO(3)$ anti-symmetric tensor $\epsilon_{ijk}$ raised by $\delta^{ja}$ that mixes the space index and the internal index.
We choose the particular scaling solution $\phi = 1/z$ with the three dimensional Euclidean symmetry. The corresponding Yang-Mills field strength $G_{\mu\nu}^a = \partial_\mu A^a_\nu - \partial_\nu A_\mu^a + g_{\mathrm{YM}} \epsilon^{a}_{\ bc} A_\mu^b A_\nu^c$ has the component
\begin{align}
g_{\mathrm{YM}} E_{i}^a &= g_{\mathrm{YM}} G_{z i}^a = \pm \frac{\delta_{i}^a}{z^2} \cr
g_{\mathrm{YM}} B_{i}^a &= -\frac{1}{2} g_{\mathrm{YM}} G_{mn}^a \epsilon_{i}^{\ mn} = - \frac{\delta_{i}^a}{z^2}
\end{align}

For later purposes, we will put the Yang-Mills coupling constant $g_{\mathrm{YM}}=2m$. The solution is (anti-)self-dual (i.e. $E_{i}^a = \pm B_{i}^a$) so that it is automatically a solution of the Einstein-Yang-Mills theory with cosmological constant due to the Weyl invariance of the Yang-Mills action as well as vanishing energy-momentum tensor for the (anti-)self-dual field strength:
\begin{align}
T_{\mu\nu}^{\mathrm{YM}}  = 2G^a_{\mu\rho}G^{a\ \rho}_{\nu}-\frac{\delta_{\mu\nu} }{2}G^{a}_{\rho\sigma}G^{a\rho\sigma} = 0   \ .
\end{align}

The metric part of the solution is  given by the Euclidean $AdS_4$ metric: 
\begin{align}
 4m^2ds^2 = \frac{dz^2 + \delta_{ij}dx^i dx^j}{z^2} \  \label{ads}
\end{align}
in the Poincar\'e coordinate.
The $SO(1,4)$ isometry of the AdS space is broken down to the Euclidean isometry + scaling isometry ($z \to \lambda z, x^i \to \lambda x^i$) without special conformal isometry due to the non-trivial Yang-Mills field configuration although the metric itself is still invariant.

In this paper, we are going to uplift the solution in eleven dimensional Euclidean supergravity \cite{Pope:1985bu}\cite{Cvetic:1999au}. The eleven-dimensional equations of motion for bosonic fields (in the notation of \cite{Pope:1985bu}) are
\begin{align}
R_{\bar{A}\bar{B}} = \frac{1}{3} \left(F_{\bar{A}\bar{C}\bar{D}\bar{E}} F_{\bar{B}}^{\ \bar{C}\bar{D}\bar{E}}-\frac{1}{12}F^2 \delta_{\bar{A}\bar{B}} \right) \cr
\nabla_{\bar{A}} F^{\bar{A}}_{\ \bar{B}\bar{C}\bar{D}} = -\frac{i}{576}\epsilon_{\bar{B}\bar{C}\bar{D}}^{\ \ \ \ \ \bar{E}_1\cdots \bar{E}_8} F_{\bar{E}_1 \cdots \bar{E}_4} F_{\bar{E}_5 \cdots \bar{E}_8}  \ , 
\end{align}
where $ F_{MNPQ} = 4 \partial_{[M} C_{NPQ]}$ is the four-form field strength. Following \cite{Pope:1985bu},  we use $\bar{A},\bar{B},\bar{C}, \cdots $ as eleven dimensional tangent space indices while ${M,N,P, \cdots}$ as coordinate indices.
Note that there is an imaginary unit $i$ in the Chern-Simons term compared with \cite{Pope:1985bu} because we are working in the Euclidean supergravity.

The Euclidean supergravity admits the Freund-Rubin solution of $AdS_4 \times S^7$ with the pure imaginary four-form flux
\begin{align}
F_{\bar{\alpha} \bar{\beta} \bar{\gamma} \bar{\delta}} = 3i m \epsilon_{\bar{\alpha} \bar{\beta} \bar{\gamma} \bar{\delta}} \ 
\end{align}
with a real constant $m$ ($\bar{\alpha},\bar{\beta},\cdots = 1,2,3,4$ are tangent space indices for the $AdS_4$). The background is dual to the three-dimensional conformal field theories with $\mathcal{N}=8$ supersymmetry with the Euclidean signature. 

It is not that odd to have an imaginary unit here because the Freund-Rubin background is the near horizon limit of the electrically charged black brane in eleven  dimensions, and the electric potential naturally acquires the imaginary unit after the analytic continuation to the Euclidean (super)gravity.
A more technical reason why we needed the imaginary unit is as follows. While the equations of motions are covariant and seem ignorant about the signature, whenever we rewrite the square of the Levi-Civita tensor as a combination of metrics, there is a sign difference between the Euclidean signature and the Lorentzian signature. The imaginary unit in the four-form flux is needed to cancel this factor, for example, in the energy-momentum tensor while keeping the appearance of the Einstein equation. The same sign issue will appear whenever we deal with the Levi-Civita tensor in the following.

In \cite{Pope:1985bu}, it was shown that the Einstein-Yang-Mills theory with $SU(2)$ gauge group in four-dimension is regarded as a consistent truncation of the eleven dimensional supergravity on $AdS_4 \times S^7$. The argument was given in the Lorentzian signature, but we are going to use the Euclidean version of the consistent truncation, in which there are a couple of difference in sign as well as the imaginary unit.  

In the consistent truncation, the eleven dimensional metric is given by
\begin{align}
ds^2 = g_{\mu\nu}(x) dx^\mu dx^\nu + g_{mn}(y) (dy^m - K^{am}(y) A_\mu^a(x) dx^\mu) (dy^n-K^{bn}(y) A_\nu^b dx^\nu) \ ,  \label{metric}
\end{align}
where $g_{\mu\nu}(x) $ is the metric for the Euclidean $AdS_4$ with the local (Poincar\'e) coordinates $x^\mu = (z,x^1,x^2,x^3)$ and  $g_{mn}(y)$ is the metric for the round seven sphere. A Killing vector $K^{a}= K^{am} \frac{\partial}{\partial y^m}$ forms a $SU(2)$ subgroup of the $SO(8)$ isometry of the seven sphere, satisfying the algebra
\begin{align}
[K^a, K^b] = -2m \epsilon^{abc} K^c \ .
\end{align} 
Geometrically, choosing the particular $SU(2)$ Killing vector $K^a$ means that we treat $S^7$ as an $SU(2)$ bundle over $S^4$.

Similarly the four-form flux is given by
\begin{align}
F_{\bar{\alpha} \bar{\beta} \bar{\gamma} \bar{\delta}} = 3im \epsilon_{\bar{\alpha} \bar{\beta} \bar{\gamma} \bar{\delta}} \cr
F_{\bar{\alpha} \bar{\beta} \bar{c}\bar{d}} = -\frac{i}{2} \tilde{G}^a_{\bar{\alpha} \bar{\beta}} M^a_{\bar{c}\bar{d}} \ , \label{flux}
\end{align}
and other components are all zero. Here $\bar{a},\bar{b},\bar{c},\cdots$ are tangent index for the internal space and $\tilde{G}^a_{\bar{\alpha}\bar{\beta}} = \frac{1}{2}\epsilon_{\bar{\alpha}\bar{\beta}\bar{\gamma} \bar{\delta}}G^{a\bar{\gamma}\bar{\delta}}$ and $M^a_{\bar{c}\bar{d}} = -m^{-1} \nabla_{[\bar{c}} K_{\bar{d}]}^a$.  

As is the case in the Lorentzian signature, but noting extra minus signs when we rewrite the square of the Levi-Civita tensor as a combination of the metric, the direct computations show that if $g_{\mu\nu}(x)$ and $G_{\mu\nu}^a(x)$ satisfy the four-dimensional Euclidean Einstein-Yang-Mills equation
\begin{align}
R_{\bar{\alpha}\bar{\beta}} - \frac{R}{2}\delta_{\bar{\alpha}\bar{\beta}} -12m^2 \delta_{\bar{\alpha}\bar{\beta}} &= 2(G^a_{\bar{\alpha}\bar{\gamma}} G^{a \ \bar{\gamma}}_{\bar{\beta}}- \frac{1}{4}G^{a}_{\bar{\gamma}\bar{\delta}} G^{a\bar{\gamma}\bar{\delta}}\delta_{\bar{\alpha}\bar{\beta}}) \ \cr
D^{\bar{\beta}}G^{a}_{\bar{\alpha}\bar{\beta}} &= 0  
\end{align}
with $g_{\mathrm{YM}} = 2m$, then the ansatz \eqref{metric} and \eqref{flux} solves the eleven-dimensional equations of motion as well as the Bianchi identity for the four-form flux. 

For our purpose, we take the particular solution \eqref{phi} and \eqref{ads} for the four-dimensional (Euclidean) Einstein-Yang-Mills equation with the negative cosmological constant. Then the uplifted eleven dimensional geometry is given by the fibration of $AdS_4$ over $S_7$, in which $AdS_4$ is rotated in the same way as we move along the $SU(2)$ bundle over the base $S^4$ of $S^7$. The four-form flux is purely imaginary but this has been already the case in the vacuum Freund-Rubin $AdS_4\times S^7$ solution in the Euclidean signature. Note, however, that from the view point of the four-dimensional effective field theory, the dimensionally reduced Yang-Mills field is actually real.

As in the four-dimensional effective theory, this uplifted Euclidean eleven dimensional solution corresponds to the fixed point of the renormalization group flow induced by the  topologically twisted deformation via the holography. The isometry of the solution corresponds to three-dimensional Euclidean symmetry (rotation and translation) and scale symmetry as well as the remaining internal $SO(5)$ symmetry (from the unbroken isometry of the internal space), but it does not include the special conformal transformation.
Here, the parent theory without the deformation is given by the M2-brane gauge theories studied in \cite{Aharony:2008ug}. The M2-brane gauge theories admit the topological twist under the $SU(2)$ subgroup of the $SO(8)$ R-symmetry. The background  \eqref{metric} and \eqref{flux} (with \eqref{phi} and \eqref{ads} substituted) therefore holographically describes a new phase of the topologically twisted M2-brane gauge theory with scale invariance without conformal invariance. 

It is noted that in our earlier studies \cite{Nakayama:2009qu}\cite{Nakayama:2009fe}\cite{Nakayama:2010wx}\cite{Nakayama:2010zz}, we argued that scale invariant but non-conformal geometry cannot be realized in healthy gravitational theories such as the eleven dimensional supergravity. The breakthrough in this paper is that we work in the Euclidean signature, and we further allow the topological twist to realize the Euclidean symmetry. The constructed geometry here cannot be analytically continued to the Lorentzian signature without violating reality conditions on supergravity fields necessary for the unitarity.

Having established the vacuum solution, the next question is if we could find the topologically twisted holographic renormalization group flow from the Freund-Rubin $AdS_4 \times S^7$ background to our new solution, which would correspond to the flow that connects the M2-brane gauge theory and our holographically constructed scale invariant but non-conformal fixed point. While it seems difficult to perform the complete analysis in the eleven dimensional supergravity, it is easier to do this at the level of the consistent truncation to the effective four-dimensional Einstein-Yang-Mills theory.

Since the holographic renormalization group geometry
\begin{align}
ds^2 = f(z) \frac{dz^2 + \delta_{ij}dx^idx^j}{z^2} 
\end{align}
is conformally flat, the general strategy is to first solve the Yang-Mills equations in the flat four-dimensional space with the three-dimensional Euclidean symmetry. In particular, one may use the $\lambda \phi^4$ theory ansatz to see if we could find the solution which behaves $\phi \sim \phi_0 + \phi_1 z$ near $z=0$ and $\phi \sim 1/z$ as $z\to \infty$. It turns out that the most generic solutions of $\lambda \phi^4$ theory with the three-dimensional Euclidean symmetry (i.e. $\phi$ depends only on $z$) is given by the Jacobi $\mathrm{sn}$ function as
\begin{align}
\phi(z) = C (-1)^{3/4} \mathrm{sn}(\frac{C(-1)^{1/4}(z+z_0)}{\sqrt{2}} , -1) \ .
\end{align}
The two integration constant $C$ and $z_0$ can be changed by the translation and the dilatation, and  we immediately see that except for the isolated limiting solution of $\phi = 1/(z+z_0)$ (i.e. $C\to \infty$), the solution always diverges at finite $z>0$ due to the periodicity of the Jacobi $\mathrm{sn}$ function.

The only reasonable solution therefore is to take $C \to 0 $ limit with $\phi = \frac{1}{z+z_0}$ so that 
\begin{align}
A_z^a = 0 \ , \ \  g_{\mathrm{YM}} A_i^a = \delta_{i}^a\frac{1}{z+z_0}  \label{gene} \ .
\end{align}
 The energy-momentum tensor remains zero, so there is no back-reaction to the $AdS_4$ metric.
While the solution may appear to be a trivial generalization of \eqref{phi},  we see that this solution gives the exact holographic renormalization group flow from the UV fixed point at $z=0$ (i.e. a relevant deformation of the Freund-Rubin background) to the IR  scale invariant but conformal fixed point we have discussed.

The AdS/CFT interpretation tells that, in the corresponding field theory side, we have the UV $\mathcal{N}=8$ superconformal field theory deformed by the twisted scalar operator:
\begin{align}
\delta S = \int d^3x J_{i}^a \delta^{i}_a \ . 
\end{align}
Here $J_{i}^a$ is an $SU(2)$ part of the R-current that is conserved at the UV superconformal fixed point.
This deformation is what we called the universal deformation associated with the topological twist in \cite{Nakayama:2016ydc} and it is always relevant with the scaling dimension $2$ at the UV fixed point (i.e. $z=0$). The scaling dimension is consistent with the profile of the bulk gauge field $A_i^a \sim \frac{\delta_{i}^a}{z+z_0} = \delta_{i}^a(\frac{1}{z_0} -\frac{z}{z_0^2} +\cdots) $.  
Then our construction tells that this deformation induces the renormalization group flow toward a scale invariant but non-conformal fixed point in the IR (i.e. $z\to \infty$).

It is remarkable that one can construct the exact renormalization group flow for the universal topologically twisted deformations in the holographic construction. Repeating the same discussion above, one my easily uplift the solution obtained in the effective field theory \eqref{gene} to the exact classical solution of the eleven dimensional Euclidean supergravity by using the same ansatz \eqref{metric} and \eqref{flux} (with \eqref{gene} substituted instead).
The fixed point we constructed preserves (Euclidean) supersymmetry without superconformal symmetry, and it seems extremely interesting to study the properties of this new phase of M2-brane gauge theories predicted by the eleven dimensional supergravity.

To conclude this paper, we would like to point out that the uplift from our solution in the effective Einstein-Yang-Mills theory may work well with less symmetric situations as long as the internal space has the $SU(2)$ isometry. The possibility of the consistent truncations depends on the number of supersymmetry and could be quite non-trivial, and it would be interesting to see if we could find the full string or M-theory background (see e.g. \cite{Gauntlett:2009zw}). In the case of tri-Sasakian compactifications, the consistent truncation was discussed in \cite{Cassani:2011fu},\footnote{The author would like to thank Nakwoo Kim for pointing out the reference.} and the construction there with our four-dimensional solution will lead to huge numbers of non-trivial scale invariant but non-conformal fixed points in three-dimensions realized in M-theory albeit available only in the Euclidean signature.

\section*{Acknowledgements}
This work is supported in part by Rikkyo University Special Fund for Research.

\end{document}